\documentclass[12pt,preprint]{aastex}
\usepackage{natbib}

\def\ga{\mathrel{\raise0.35ex\hbox{$\scriptstyle >$}\kern-0.6em
\lower0.40ex\hbox{{$\scriptstyle \sim$}}}}
\def\la{\mathrel{\raise0.35ex\hbox{$\scriptstyle <$}\kern-0.6em
\lower0.40ex\hbox{{$\scriptstyle \sim$}}}}

\def\cotwo{CO {\it J}=2--1 }

\def\coseven{CO {\it J}=7--6 }

\def\cofive{CO {\it J}=5--4 }
\def\hij{high-{\it J} }
\def\loj{low-{\it J} }

\slugcomment{Accepted for publication in ApJ}
 
\shorttitle{Cold dust and molecular gas in hyperluminous infrared quasar host galaxies}
\shortauthors{Wagg et al.}
 
\begin{document}

\title{Karl G. Jansky Very Large Array observations of cold dust and molecular gas in starbursting quasar host galaxies at $z \sim 4.5$}

\author{J.~Wagg\altaffilmark{1,2,3}, C.~L.~Carilli\altaffilmark{4,2}, M. Aravena\altaffilmark{3,5}, P. Cox\altaffilmark{6}, L.~Lentati\altaffilmark{2}, R.~Maiolino\altaffilmark{2}, R.~G.~McMahon\altaffilmark{7}, D.~Riechers\altaffilmark{8}, F.~Walter\altaffilmark{9}, P.~Andreani\altaffilmark{10}, R. Hills\altaffilmark{2}, and  A.~Wolfe\altaffilmark{11}}

\email{j.wagg@skatelescope.org}
\altaffiltext{1}{Square Kilometre Array Organisation, Lower Withington, UK}
\altaffiltext{2}{Cavendish Laboratory, University of Cambridge, Cambridge UK}
\altaffiltext{3}{European Southern Observatory, Casilla 19001, Santiago, Chile}
\altaffiltext{4}{National Radio Astronomy Observatory, Socorro, USA}
\altaffiltext{5}{N\'{u}cleo de Astronom\'{\i}a, Facultad de Ingenier\'{\i}a, Universidad Diego Portales, Santiago, Chile}
\altaffiltext{6}{Joint ALMA Observatory, Santiago, Chile}
\altaffiltext{7}{Institute of Astronomy, University of Cambridge, Cambridge, UK}
\altaffiltext{8}{Cornell University, Ithaca, USA}
\altaffiltext{9}{Max-Planck Institute for Astronomy, Heidelberg, Germany}
\altaffiltext{10}{European Southern Observatory, Garching, Germany}
\altaffiltext{11}{Department of Physics and Center for Astrophysics and Space Sciences, UCSD, USA}

\begin{abstract}
\noindent 

We present Karl G. Jansky Very Large Array (VLA) observations of 44~GHz continuum and \cotwo line emission in  BR1202-0725 at $z = 4.7$ (a starburst galaxy and quasar pair) and BRI1335-0417 at $z=4.4$ (also hosting a quasar). 
With the full 8~GHz bandwidth capabilities of the upgraded VLA, we study the (rest-frame) 250~GHz thermal dust continuum emission for the first time along with the cold molecular gas traced by the \loj CO line emission.
The measured \cotwo line luminosities of BR1202-0725 are $L'_{CO} = (8.7\pm0.8)\times 10^{10}$~K~km~s$^{-1}$~pc$^{2}$ and $L'_{CO} = (6.0\pm0.5)\times 10^{10}$~K~km~s$^{-1}$~pc$^{2}$ for the submm galaxy (SMG) and quasar, which are equal to previous measurements of the \cofive line luminosities implying thermalized line emission and we estimate a combined cold molecular gas mass of $\sim$9$\times$10$^{10}$~M$_{\odot}$. In BRI1335-0417 we measure $L'_{CO} = (7.3\pm0.6)\times 10^{10}$~K~km~s$^{-1}$~pc$^2$. 
We detect continuum emission in the SMG BR1202-0725 North ($S_{44GHz} = 51\pm6$~$\mu$Jy), while the quasar is detected with $S_{44GHz} = 24\pm6$~$\mu$Jy and in BRI1335-0417 we measure $S_{44GHz} = 40\pm 7$~$\mu$Jy. 
 Combining our continuum observations with previous data at (rest-frame) far-infrared and cm-wavelengths, we fit three component models in order to estimate the star-formation rates.  This spectral energy distribution fitting suggests that the dominant contribution to the observed 44~GHz continuum is thermal dust emission, while either thermal free-free or synchrotron emission contributes less than 30\%.  

\end{abstract}


\section{Introduction}

The high-redshift formation of some of the most massive present-day galaxies is often accompanied by episodes of extreme far-infared (FIR) luminosity, as high-lighted by the phenomena of submm/mm bright quasar host galaxies and starbursting submm galaxies (SMGs). The interpretation of dust-heated star-formation to explain the rest-frame FIR continuum properties of these populations is supported by the detection of redshifted [CII] line emission (e.g. Maiolino et al.\ 2005; Stacey et al.\ 2010; Venemans et al.\ 2012; Wang et al.\ 2013), as this is typically the strongest cooling line in the FIR spectrum of nearby galaxies (e.g. Stacey et al.\ 1991).  Fueling the star-formation and active galactic nucleus (AGN) activity requires significant molecular gas reservoirs, most efficiently studied through observations of redshifted CO line emission (see Carilli \& Walter 2013 for a recent review). Together, these line and continuum observations can probe the physical conditions and kinematic properties of the interstellar medium in galaxies over much of the history of the Universe. 

Recent studies of gas and dust in high-redshift galaxies have used photoionization models along with measured values of the relative intensity between the FIR continuum, low-\textit{J} CO and [CII] line emission to constrain the ionization rate and density of the interstellar medium (e.g. Stacey et al.\ 2010; Ivison et al.\ 2010). With new facilities like the Atacama Large Millimeter/submillimeter Array (ALMA) and the upgraded VLA it is now possible to perform similar analyses at kpc scale resolution with some of the most distant starburst galaxies and AGN, while also constraining their kinematic properties (e.g. Riechers et al.\ 2011; Ivison et al. 2011; Wang et al.\ 2013; Carilli et al.\ 2013). Furthermore, with the combined continuum sensitivity of these interferometers one can constrain the emissivity of the thermal dust continuum emission and obtain independent estimates of the star-formation rates (SFRs) through detection of thermal free-free emission.

The FIR luminous quasar host galaxies, BRI1335-0417 and BR1202-0725, are two well-studied cases of massive, unlensed galaxy formation roughly 1.5~Gyr after the big bang. BRI1335-0417 is believed to be a ``wet'' merger at $z = 4.4$ with an estimated FIR luminosity of $L_{FIR} \sim 3\times 10^{13}$~L$_{\odot}$ fueled by $\sim9 \times 10^{10}$~M$_{\odot}$ of cold molecular gas traced by CO line emission (Storrie-Lombardi et al.\ 1996; Ohta et al.\ 1996; Guilloteau et al.\ 1997; Benford et al.\ 1999; Carilli et al.\ 1999, 2002; Riechers et al.\ 2008). Similarly, BR1202-0725 at $z = 4.7$ is a quasar host galaxy discovered in the APM survey (Irwin et al.\ 1991) and has been shown to be comprised of two FIR luminous radio sources separated by $\sim$3.8$''$ (Omont et al. 1996; Yun et al.\ 2000). Both the quasar host galaxy BR1202-0725 South and the optically faint SMG companion to the North have FIR luminosities in excess of 10$^{13}$~L$_{\odot}$ and a combined cold molecular gas mass of $\sim$10$^{11}$~M$_{\odot}$ (Carilli et al.\ 2002; Iono et al.\ 2006; Riechers et al.\ 2006). Within the same region, Hu et al.\ (1996) identified two fainter Ly$\alpha$ companions, both of which are thought to exhibit faint 157.7$\mu$m [CII] line emission (Carilli et al.\ 2013). Submm interferometric observations of [CII] and dust continuum emission with ALMA show that the Lyman-$\alpha$ companion to the SW of the quasar is also luminous in the FIR (Wagg et al.\ 2012; R.~Williams et al.\ \textit{in prep.}). 
It is likely that BR1202-0725 is at an early stage of a merger between multiple gas-rich systems (Salom\'e et al.\ 2012).

In this work we present VLA observations of rest-frame 250~GHz continuum and redshifted \cotwo line emission in the quasar host galaxies, BR1202-0725 and BRI1335-0417. We adopt a cosmological model with $(\Omega_\Lambda, \Omega_m, h) = (0.73, 0.27, 0.71)$ (Spergel et al.\ 2007).

\section{Observations and data reduction}

\begin{figure}
\epsscale{0.5}
\plotone{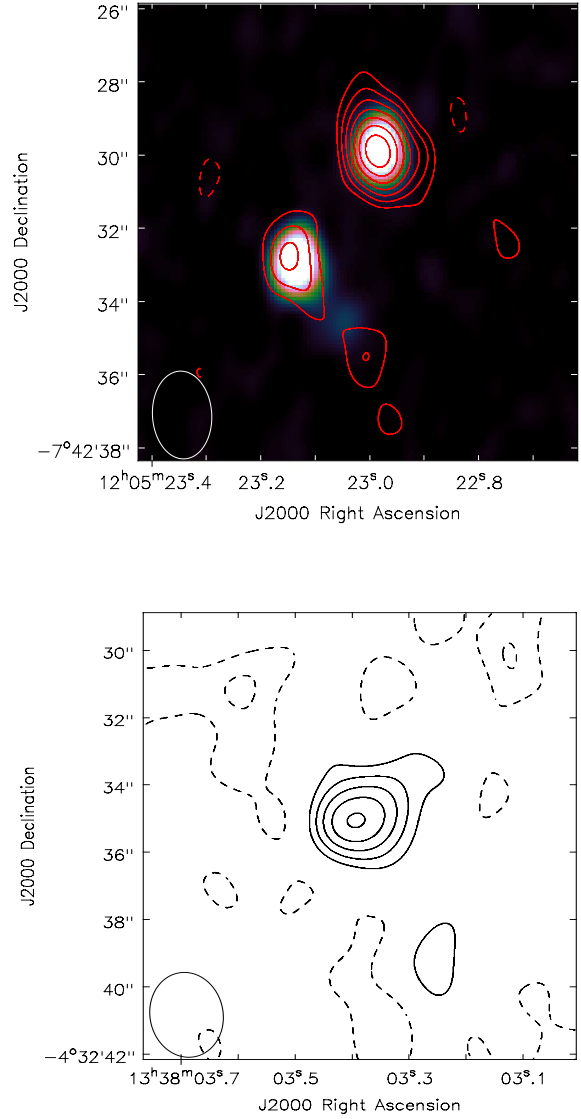}

\caption{
\textit{top:} Contours of the VLA 44~GHz continuum emission from BR1202-0725 overlaid on the image of the 340~GHz continuum emission observed with ALMA (Wagg et al.\ 2012). The rms in the 340~GHz continuum image is 0.4~mJy~beam$^{-1}$ and the synthesized beamsize is $1.30'' \times 0.86''$ (not shown). Contour intervals are (-2, 2, 3, 4, 5, 6 , 7 and 8)$\times 5.9~\mu$Jy per beam and the synthesized beamsize is $2.40'' \times 1.60''$ (PA = 3.4$^\circ$; bottom left). 
\textit{bottom:} Contour map of the 44~GHz continuum emission in BR1335-0417 at $z = 4.4$. The beamsize is $2.53'' \times 1.92''$ (PA = 6.4$^\circ$) and the contour levels are (-3, -2, 2, 3, 4, 5 and 6)$\times 6.5 \mu$Jy per beam. 
}
\end{figure}

Observations were carried out with 27 elements of the VLA on four dates in January and February of 2013, while the array was in the most compact D-configuration. The rest frequency of the CO~\textit{J}=2-1 line is 230.538~GHz, so that at the redshifts of our targets this line can be observed at $\sim$40.5~GHz for the two most luminous galaxies in the BR1202-0725 system, and at 42.6~GHz in BRI1335-0417.  
 Each field was observed for a total of 8~hours including overheads associated with complex gain, bandpass and flux density calibration. In the case of each target field, 3C286 was used to calibrate the flux density and bandpass shape, while J1229+0203 and J1354-0206 were used to calibrate the complex gains for BR1202-0725 and BRI1335-0417, respectively. Using the recently commissioned 8~GHz bandwidth correlator mode with  the Q-band receivers our observations covered 40 to 48~GHz with a spectral resolution of 2~MHz. The frequency coverage in this mode is not contiguous, and small 8~MHz gaps between sub-bands mean that two slightly offset tunings were used to fully sample the entire range. 

The data reduction was performed using the \textit{CASA} software package\footnote{http://casa.nrao.edu}. Standard calibration steps are carried out using a new set of \textit{CASA} pipeline reduction procedures developed by NRAO staff. Including only the spectral windows without line emission, we then use natural weighting and multi-frequency synthesis imaging to generate the continuum images, achieving sensitivities of 5.9$\mu$Jy and 6.5$\mu$Jy~beam$^{-1}$ for BR1202-0725 and BR1335-0417, respectively. For the spectral line cubes, we also used natural weighting in the imaging and the synthesized beamsizes are $2.55'' \times 1.95''$ (PA = 2.3$^\circ$) and $2.57'' \times 1.75''$ (PA = 7.4$^\circ$) in the two images. The spectral line maps are resampled to 8~MHz resolution, corresponding to velocity channel widths of 59 and 56~km~s$^{-1}$ at the redshifted frequencies of the CO~\textit{J}=2-1 line in BR1202-0725 and BRI1335-0417, while the rms per channel is 160 and 170$\mu$Jy, respectively.

\section{Results}

\begin{table*}[]
\centering
\caption{Observed line and continuum properties of BR1202-0725 and BRI1335-0417.}
\begin{center}
\begin{tabular}{l c c c c c } 
\hline \hline
Source  & $S_{44GHz}$ & $z _{\rm [CO]}$$^a$ &  $\Delta V_{FWHM}$$^a$     & $SdV$ & $L'_{\rm [CO~\textit{J=}2-1]}$  \\ 
        &    [$\mu$Jy]  &   &   [km~s$^{-1}$]  &   [Jy~km~s$^{-1}$] &  $\times10^{10}$ [K~km~s$^{-1}$~pc$^{2}$] \\ 
\hline 
BR1202-0725 North  &  51$\pm$6  &   4.692    &   1108$\pm$60  &   0.42$\pm$0.04   &  8.7$\pm$0.8  \\ 
BR1202-0725 South  &  24$\pm$6  &   4.694    &    352$\pm$18  &   0.29$\pm$0.02  &   6.0$\pm$0.5  \\ 
BRI1335-0417   &   40$\pm$7  &    4.406          &    322$\pm$13   &    0.38$\pm$0.03  & 7.3$\pm$0.6  \\ 
\hline
\end{tabular}
\vskip 0.1in
\noindent $^a${} The \cotwo redshifts and line widths ($\Delta V_{FWHM}$) are determined from the best-fit parameters of a Gaussian fit to the spectra. \\
\label{ref:table1}
\end{center}
\end{table*}
\vskip 0.1in

Figure~1 shows the 44~GHz continuum images of the two target fields, where we also overlay contours of the 340~GHz continuum observed in BR1202-0725 with ALMA. 
The two most luminous submm sources in the BR1202-0725 field are detected at 44~GHz, with peak flux densities of $53\pm 6\mu$Jy and $24\pm 6 \mu$Jy, for the Northern and Southern components. The integrated flux densities are similar to the peak values for both components so they are unresolved. BRI1335-0417 is also detected with a peak flux density of $40\pm 7 \mu$Jy, which is in agreement with the prediction by Carilli et al.\ (1999) who assume a thermal dust emissivity index, $\beta = 1.5$. 
Future higher spatial resolution continuum observations will determine the size of this emission region.

From the spectral line data cubes we extract spectra of the \cotwo line emission at the positions of our three primary targets (Figure~2). For comparison, we show the [CII] line profiles for the BR1202-0725 SMG and quasar host galaxy measured from recent ALMA commissioning observations (Wagg et al.\ 2012; Carilli et al.\ 2013; Lentati et al.\ 2013). Table~1 gives the redshift, line widths and integrated intensities of the \cotwo line emission observed here with the VLA, which have been calculated assuming the data can be decribed by a single Gaussian line profile. However, based on the [CII] line observations and the \cofive and \coseven line emission observed with the Plateau de Bure Interferometer (PdBI) by Salom\'e et al.\ (2012), the line profiles of BR1202-0725 North appear to be best described by two Gaussian components. The \cotwo redshift and line width for the quasar BR1202-0725 South are in excellent agreement with previous observations of [CII] and CO line emission (Omont et al.\ 1996; Carilli et al.\ 2002; Wagg et al.\ 2012), however these data hint at the presence of broad wings in both sources, and more sensitive observations are needed to confirm this.

In the case of BRI1335-0417, the \cotwo line profile measured here is slightly narrower than that of the previous [CII] and \cofive lines detected ($\sim430$~km~s$^{-1}$; Guilloteau et al.\ 1997; Wagg et al.\ 2010), and also blueshifted by $\sim$50~km~s$^{-1}$. The integrated intensity of the \cotwo line emission measured here (0.38$\pm$0.03~Jy~km~s$^{-1}$) is in good agreement with the previous observations reported by Riechers et al.\ (2008), who measure 0.43$\pm$0.02~Jy~km~s$^{-1}$ using the old VLA correlator. This implies that the narrower bandwidth of past observations was wide enough to encompass the velocity width of the line, suggesting that the differences in the linewidths can be attributed to the low signal-to-noise of the [CII] and \cofive spectra. 

\begin{figure*}
\epsscale{0.7}

\plotone{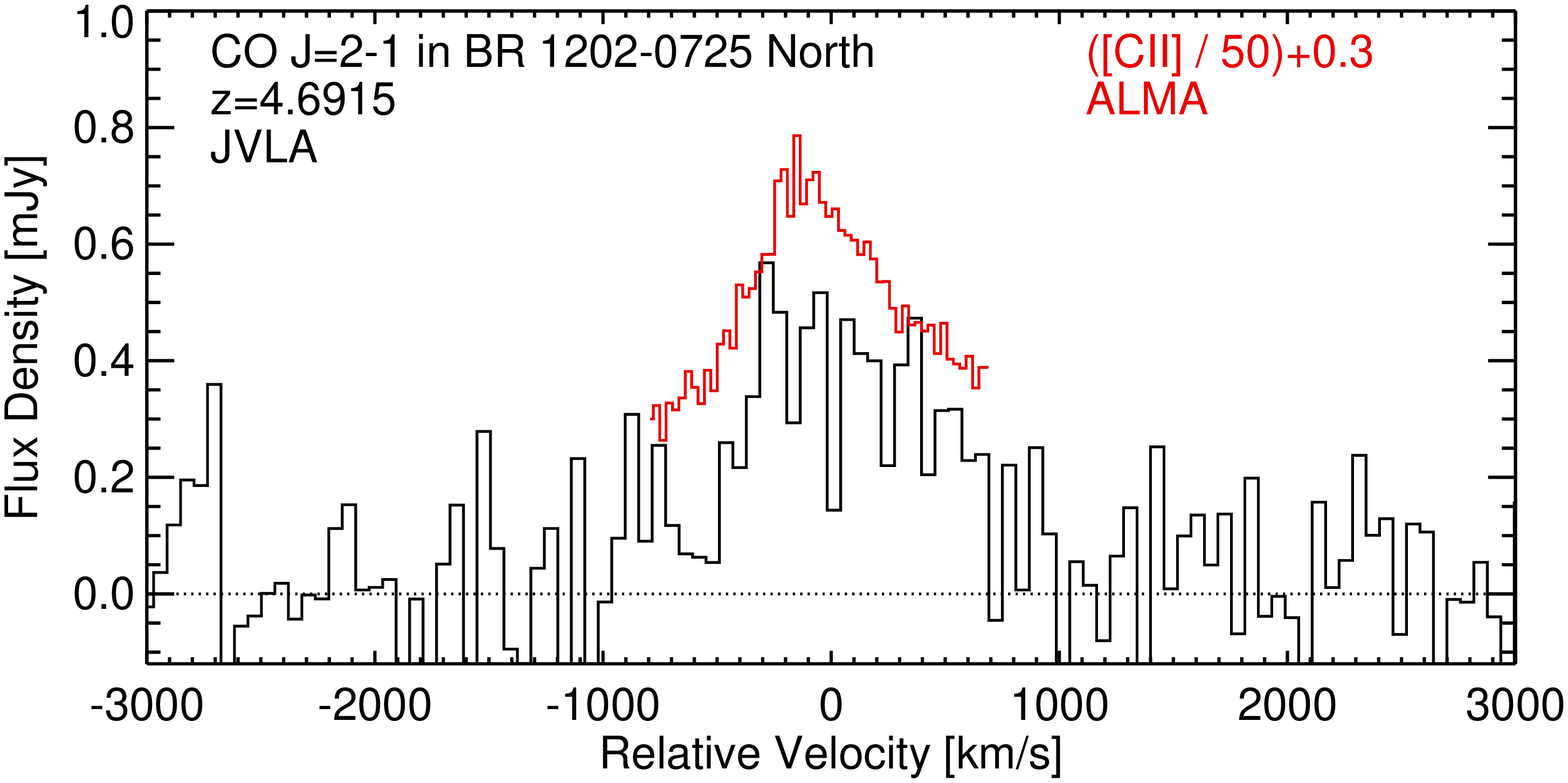}

\plotone{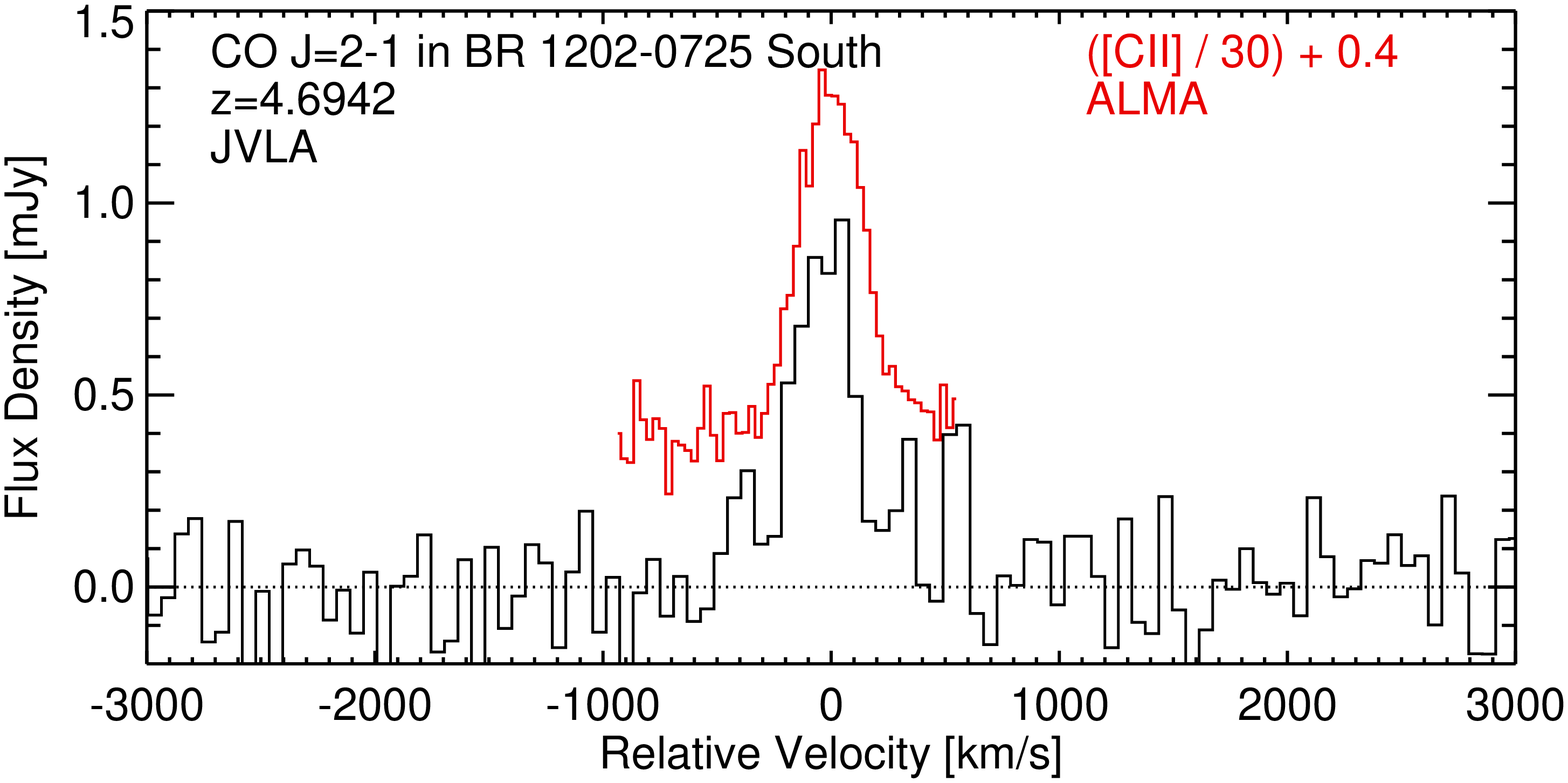}

\plotone{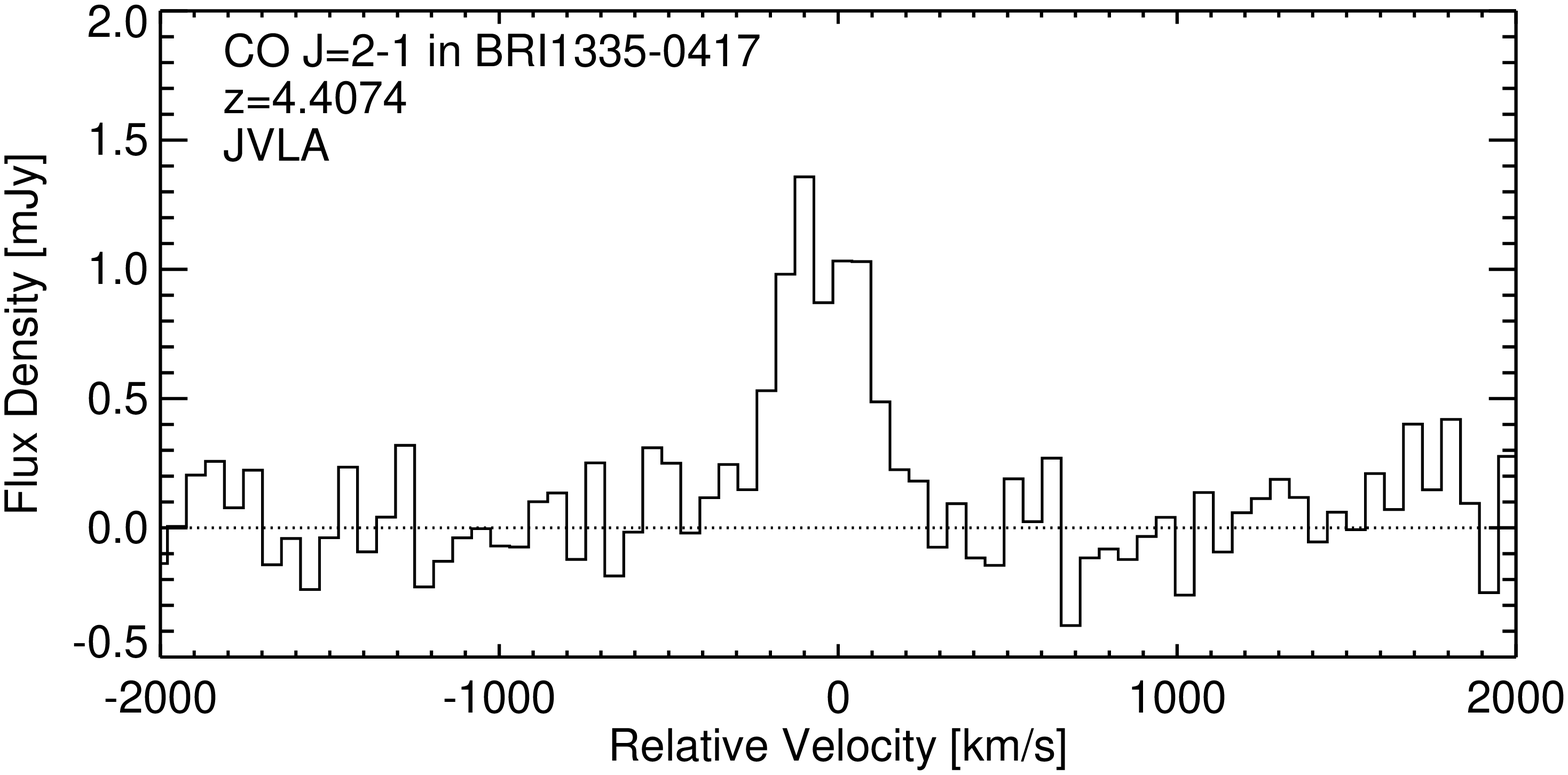}

\caption{\textit{top:} Spectra of \cotwo line emission in BR1202-0725 North and South compared with  the continuum subtracted [CII] line profiles measured by ALMA. The continuum subtracted [CII] line emission has been scaled down and offset from zero for comparison with the CO. 
The velocity offsets are relative to z=4.6915 and 4.6942, and the rms per 59~km~s$^{-1}$ channel is 160~$\mu$Jy. \textit{bottom:} The \cotwo line emission detected in BRI1335-0417 at $z  = 4.4065$. The rms per 56~km~s$^{-1}$ channel is 170~$\mu$Jy.  
}
\end{figure*}

\section{Analysis}

\subsection{Continuum emission}

We use archival \textit{Herschel} SPIRE submm-wavelength and lower frequency, 1.4 and 5~GHz VLA observations of our targets (project AC878) to better sample their (rest-frame) FIR through cm-wavelength spectral energy distributions. 
  For BR1202-0725 the FIR spectral energy distribution has been observed with ALMA, PdBI and \textit{Herschel} SPIRE. From the level 2 pipeline processed SPIRE maps we extract flux densities at the pixels corresponding to the 44~GHz peak position of BRI1335-0417, measuring, $S_{250\mu m} = 33\pm6$~mJy, $S_{350\mu m} = 41\pm5$~mJy, and $S_{500\mu m} = 44\pm6$~mJy, while in the case of BR1202-0725 the total 350~$\mu$m flux density of both components measured in the SPIRE beam is $S_{350\mu m} = 78\pm 6$~mJy. Figure~3 shows the observed spectral energy distributions for BR1202-0725 and BRI1335-0417. 

In order to model the spectral energy distribution we assume the data can be described by a power-law synchrotron component dominating at frequencies below $\sim$30~GHz, a thermal free-free component with a power-law index of $\alpha_{ff} =-0.1$, and a single temperature greybody component which is described by a dust temperature, $T_d$ and emissivity index, $\beta$. Following Yun \& Carilli (2002), we use the linear relationships between the total SFR and the three components describing the radio-to-FIR spectral energy distribution (Condon 1992) to 
obtain simultaneous constraints on all parameters in the model. The parameters we fit are therefore synchrotron spectral index, $\nu^{-\alpha}$, a scaling factor for the non-thermal synchrotron emission ($f_{nth}$; see equation (13) of Yun \& Carilli 2002), and finally the SFR, $T_d$ and $\beta$. 
Our fitting method is based on the principles of Bayesian inference, which provides a consistent approach to the estimation
of a set of parameters $\Theta$ in a model or hypothesis $H$ given the data, $D$.  Bayes' theorem states that:

\begin{equation}
\mathrm{Pr}(\Theta \mid D, H) = \frac{\mathrm{Pr}(D\mid \Theta,
H)\mathrm{Pr}(\Theta \mid H)}{\mathrm{Pr}(D \mid H)},
\end{equation}
where $\mathrm{Pr}(\Theta \mid D, H) \equiv \mathrm{Pr}(\Theta)$ is the posterior probability distribution of the parameters,  $\mathrm{Pr}(D\mid
\Theta, H) \equiv L(\Theta)$ is the likelihood, $\mathrm{Pr}(\Theta \mid H) \equiv \pi(\Theta)$ is the prior probability distribution, and
$\mathrm{Pr}(D \mid H) \equiv Z$ is the Bayesian evidence.

In parameter estimation, the normalizing evidence factor is usually ignored, since it is independent of the parameters $\Theta$.   Inferences
are therefore obtained by taking samples from the (unnormalised) posterior using, for example, standard Markov chain Monte Carlo (MCMC) sampling
methods. An alternative to MCMC is the nested sampling approach (Skilling 2004), a Monte-Carlo method targeted at the efficient
calculation of the evidence, that also produces posterior inferences as a by-product.  In Feroz et al.\ (2008, 2009) this nested sampling framework was developed further with the introduction of the MultiNest algorithm, which provides an efficient means of sampling from posteriors that may contain multiple modes and/or large (curving) degeneracies, and also calculates the evidence. We make use of the MultiNest algorithm to obtain our estimates of the posterior probability distributions for the spectral energy distribution parameters. 

\begin{figure*}
\epsscale{0.5}

\plotone{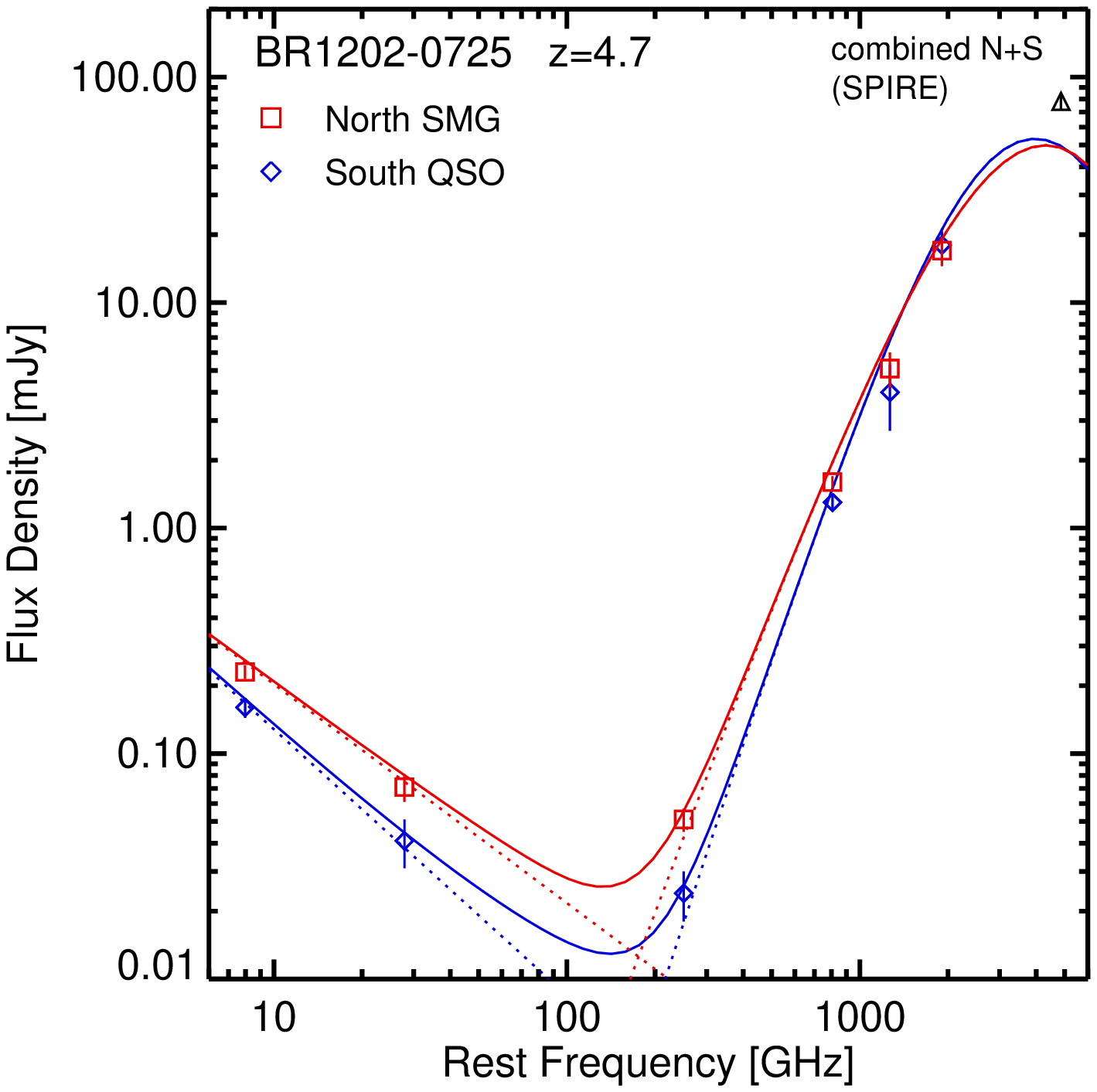}

\plotone{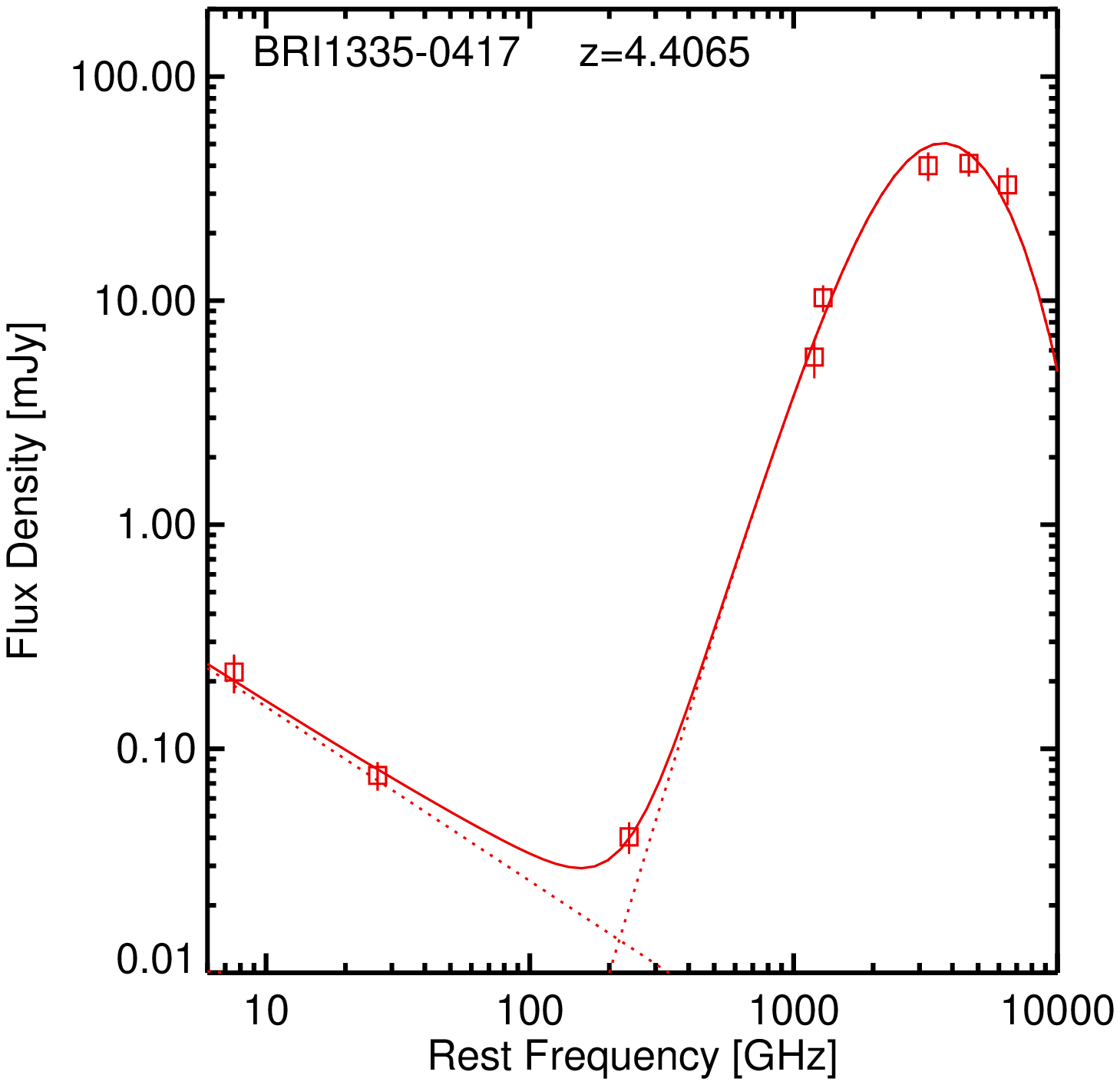}

\caption{\textit{top:} Far-infared spectral energy distributions of  BR1202-0725 North and South along with the best-fitting greybody, synchrotron and thermal free-free models described in the text and parameters given in Table~2. Observed data are from Carilli et al. (2002), this work, Salom\'e et al.\ (2012), and Wagg et al. (2012). The \textit{solid} curves show the best-fit models while the \textit{dotted} lines show the synchrotron and thermal dust components. The thermal free-free component of the model is not shown but does make a small contribution to the \textit{solid} line.  
\textit{bottom:} The observed spectral energy distribution of BRI1335-0417 plotted with the best-fitting model (\textit{solid} line) composed of the thermal dust continuum and synchrotron emission curves (\textit{dotted} lines), and only a small contribution from the thermal free-free emission. Observed data are from Carilli et al.\ (1999), this work, and Guilloteau et al.\ (1997). }
\end{figure*}

Naturally, the model parameters are correlated. For example, when fitting thermal dust continuum emission there are degeneracies between the emissivity index, $\beta$, and $T_d$ (e.g. Priddey \& McMahon 2001; Blain et al.\ 2003). These correlations can be seen in the probability density plots shown in Figure~4 and the mean of the posterior parameter values are given in Table~\ref{ref:table2}. Although we impose the prior that the total 350~$\mu$m emission in BR1202-0725 North and South should be equal to the flux density measured by SPIRE, the relative temperatures of these two components is unknown. In an attempt to overcome this, we include priors on the dust temperatures based on recent submm-wavelength observations of  high-redshift starburst galaxies ($T_d = 40$~K; e.g. Magnelli et al.\ 2012) and quasar host galaxies ($T_{d} = 50$~K; Beelen et al.\ 2006; Wang et al.\ 2008).  However, we find that the resulting fits do a poor job of reproducing the measured 44~GHz flux densities, and so we adopt the values derived in the free floating parameter analysis.

Evidence for thermal free-free emission has recently been observed in gravitationally lensed, FIR luminous starburst galaxies  (Thomson et al.\ 2012; Aravena et al.\ 2013). The models provide an estimate of the contribution from such free-free emission to the observed 44~GHz continuum emission in our targets. 
Based on the fitting analysis, the main contribution to the 44~GHz flux density in each source is thermal dust continuum emission. In the case of BRI1335-0417, that contribution is 54\%, with 30\% coming from synchrotron emission and only 16\% from thermal free-free emission. In the case of the SMG BR1202-0725 North, the thermal dust emission contribution from our model fit is 75\%, while 17\% and 8\% come from synchrotron and free-free emission, respectively. The model fit to the spectral energy distribution of the quasar BR1202-0725 South has a 67\% contribution to the 44~GHz flux density from thermal dust emission, while only 12\% comes from synchrotron emission and 21\% is found to come from free-free emission. Although it is not considered in our models, another possibility is that the 44~GHz emission arises from a flat-spectrum AGN embedded in the starburst host galaxies, however higher resolution imaging is needed before we can determine the likelihood of this.

\subsection{\cotwo Line Emission}

Following the definition of line luminosity given by Solomon et al.( 1992), we calculate the line luminosities to be $L'_{CO} = (8.7\pm0.8)\times 10^{10}$~K~km~s$^{-1}$~pc$^{2}$ and $L'_{CO} = (6.0\pm0.5)\times 10^{10}$~K~km~s$^{-1}$~pc$^{2}$ for BR1202-0725 North and South, respectively, and $L'_{CO} = (7.3\pm0.6)\times 10^{10}$~K~km~s$^{-1}$ for BRI1335-0417. For BR1202-0725, these luminosities are in excellent agreement with previous lower spectral resolution observations with the VLA (Carilli et al.\ 2002), and with the \hij CO lines measured by Omont et al.\ (1996) and Salom\'e et al.\ (2013), and supports the claim that the line emission is thermalized.  This is also the case for BRI1335-0417, for which the \cofive line luminosity,  $L'_{CO} = (8.5\pm0.9)\times 10^{10}$~K~km~s$^{-1}$~pc$^{2}$  measured by Guilloteau et al.\ (1997) is similar to what we measure in \cotwo.

ALMA observations reveal [CII] line emission in the two LAEs at $z\sim 4.7$ associated with the BR1202-0725 system (Carilli et al.\ 2013). We search the Q-band spectral line data cubes at the positions of Ly$\alpha$-1 and Ly$\alpha$-2 and the expected frequencies of redshifted \cotwo , but this emission is undetected. Assuming Gaussian line profiles we set 3-$\sigma$ upper limits to the line luminosities, $L'_{CO} < 4.8 \times 10^{9}$~K~km~s$^{-1}$~pc$^{2}$ and $L'_{CO} < 1.2 \times 10^{10}$~K~km~s$^{-1}$~pc$^{2}$ for Ly$\alpha$-1 and Ly$\alpha$-2 assuming FWHM linewidths of 56 and 338~km~s$^{-1}$. 

\begin{figure*}[ht]
\epsscale{0.6}

\plotone{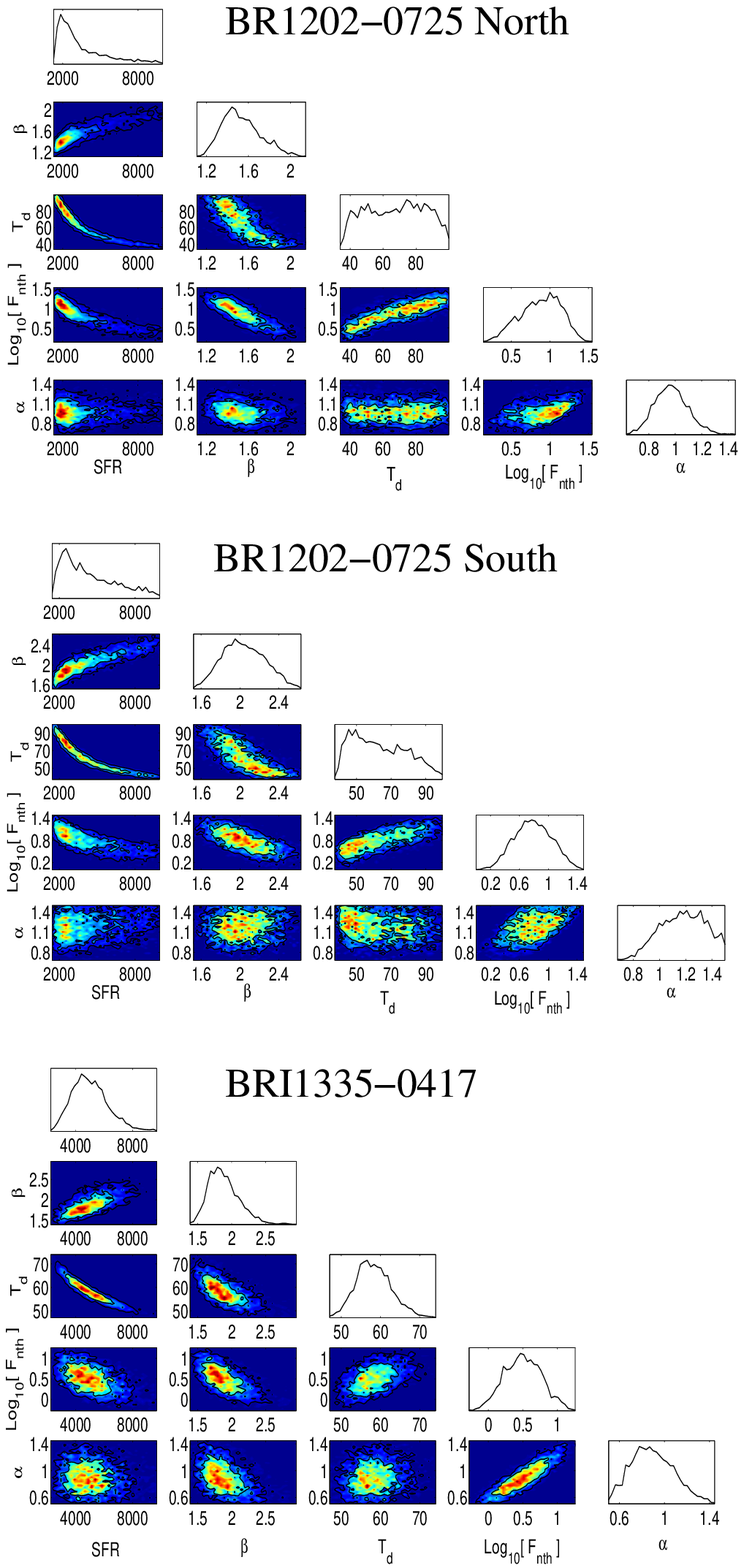}

\caption{Parameter probability density plots for our three targets resulting from the MultiNest Bayesian fitting analysis. Each plot shows how the fit parameters are correlated when fitting the model described in the text to the observed data. The curves plotted at the top of each column indicate the probability distribution for the parameter in the label at the bottom. The mean of the posterior values and 1-$\sigma$ confidence intervals are given in Table~2.}
\end{figure*}

\section{Discussion}

Although the uncertainties are large, the star-formation rates estimated for the three targets are similar to previous estimates derived from the FIR luminosities (e.g. Wagg et al.\ 2010; Salom\'e et al.\ 2012; Carniani et al.\ 2013). 
  The FIR emission model fit to the SMG BR1202-0725 North favours an emissivity index of $\beta \sim 1.5$, while the quasars BR1202-0725 South and BRI1335-0417 appear to have values closer to $\beta \sim 2.0$. Both of these values are similar to what has been measured in 60$\mu$m selected galaxies in the nearby Universe (e.g. Dunne et al.\ 2000). More recent studies of the full FIR through submm-wavelength emission in normal star-forming galaxies from the KINGFISH sample show that nearby galaxies can exhibit a broad range in emissivity indices, $\beta \sim$0.8--2.5 (Galametz et al.\ 2012). Our measured values are therefore not unusual. If free-free emission does not contribute significantly to the 44~GHz emission observed in BR1202-0725 North, then these data suggest that the cold dust mass is a factor of $\sim$2.5$\times$ higher than that in the quasar BR1202-0725 South. 

In all three sources, the contribution from thermal free-free emission to the rest-frame 250~GHz continuum is $\la$20\%. Free-free emission typically contributes less than 10\% of the 230~GHz continuum emission observed in normal, nearby galaxies (e.g. Albrecht et al.\ 2007), and so our results are consistent with what is observed in these more quiescent star-forming galaxies. Strong free-free emission has been detected in lensed SMGs (Aravena et al.\ 2013), and the SFRs estimated are in good agreement with those estimated from the FIR luminosities. 

Our measured \cotwo line luminosities are in good agreement with previous observations of both the low and \hij CO emission observed in these objects (Guilloteau et al.\ 1997; Carilli et al.\ 2002; Riechers et al.\ 2006, 2008). The luminosities imply cold molecular gas masses $\sim$10$^{11}$~M$_{\odot}$ assuming the recently measured value $\alpha_{CO} = 0.6~$M$_{\odot}$~[K~km~s$^{-1}$~pc$^{2}$]$^{-1}$  for the CO-to-H$_2$ conversion factor in ULIRGs (Papadopoulos et al.\ 2012), similar to the original value of Downes \& Solomon (1998). However, we note that this factor can be uncertain by $\sim$2. A recent analysis of the [CII] line emission observed in BR1202-0725 infers dynamical masses for the quasar and SMG based on the kinematic properties of the gas (Carniani et al.\ 2013). They calculate dynamical masses of $M_{dyn} \sim 4\times 10^{10}$~M$_{\odot}$ and $\sim 6\times 10^{10}$~M$_{\odot}$, for the quasar and SMG, implying very high molecular gas mass fractions of $\sim 80-90$\%. Such high gas fractions are inferred for some local ultraluminous infared galaxies (e.g. Papadopoulos et al.\ 2012). Planned sub-arcsecond imaging of this cold molecular gas in these targets would be complemented by higher resolution imaging of the (rest-frame) FIR continuum and [CII] line emission with ALMA in order to infer the ionizing radiation field and gas density on kpc scales.

\section{Summary}
\label{sec:sum}

We present VLA observations of redshifted \cotwo line and thermal dust continuum emission in high-redshift quasar host galaxies, observing BR1202-0725 and BRI1335-0417. The luminosities measured in the \cotwo line detected in all three targets are similar to previous observations of \cofive line emission, implying that the gas kinetic temperature is close to the excitation temperature. The observations reveal strong (rest-frame) 250~GHz continuum emission associated with all three FIR luminous galaxies, with stronger emission observed in the two starburst galaxies than the quasar BR1202-0725 South. Although synchrotron and thermal free-free likely contribute to some of this emission, we intrepret the remainder as due to thermal dust emission. Future observations at frequencies between 10 and 40~GHz will allow us to better constrain the relative importance of the three processes.

\acknowledgements We thank Maud Galametz, Rob Kennicutt, Thomas Greve and Padelis Papadopoulos for helpful discussions,  and the anonymous referee for their comments on the submitted manuscript.
This work was co-funded under the Marie Curie Actions of the European Commission (FP7-COFUND). We thank all those involved in the VLA project for making these observations possible (project code 13A-012). This paper makes use of the following ALMA data: ADS/JAO.ALMA\#2011.0.00006.SV. ALMA is a partnership of ESO (representing its member states), NSF (USA) and NINS (Japan), together with NRC (Canada) and NSC and ASIAA 
(Taiwan), in cooperation with the Republic of Chile. The Joint ALMA 
Observatory is operated by ESO, AUI/NRAO and NAOJ. The National Radio Astronomy Observatory is a facility of the National Science Foundation operated under cooperative agreement by Associated Universities, Inc.

\begin{table*}[h]
\centering
\caption{Best-fit spectral energy distribution model parameters for BR1202-0725 and BRI1335-0417.}
\begin{center}
\begin{tabular}{lccccc}
\hline \hline
Source                        & SFR    & $\alpha$  & $\log f_{nth}$  & $T_d$ [K]  & $\beta$  \\
                               & [M$_{\odot}$~yr$^{-1}$] &        &    & [K]   &     \\
\hline                         
BR1202-0725 North  &  3508$\pm$1990   &  0.97$\pm$0.13  & 0.88$\pm$0.26   &  67.8$\pm$17.5   & 1.54$\pm$0.19  \\
BR1202-0725 South  &  4411$\pm$2141   &  1.17$\pm$0.17  & 0.79$\pm$0.26   &  63.7$\pm$15.6  &  2.04$\pm$0.22   \\
BRI1335-0417           &  5040$\pm$1304   &  0.91$\pm$0.18  & 0.52$\pm$0.27   &  58.3$\pm$4.5   & 1.89$\pm$0.23   \\
\hline
\end{tabular}
\vskip 0.1in
\noindent {} Values quoted are the mean and 1-$\sigma$ uncertainty.
\label{ref:table2}
\end{center}
\end{table*}
\vskip 0.1in


\begin{thebibliography}{}

\bibitem[Albrecht et al.(2007)]{2007A&A...462..575A} Albrecht, M., Kr{\"u}gel, E., \& Chini, R.\ 2007, \aap, 462, 575 
\bibitem[Aravena et al.(2013)]{2013MNRAS.433..498A} Aravena, M., Murphy, E.~J., Aguirre, J.~E., et al.\ 2013, MNRAS, 433, 498 
\bibitem[Beelen et al.(2006)]{2006ApJ...642..694B} Beelen, A., Cox, P., Benford, D.~J., et al.\ 2006, \apj, 642, 694 
\bibitem{} Benford, D.~J., Cox, P., Omont, A., Phillips, T.~G., \& McMahon, R.~G.\ 1999, \apjl, 518, L65 
\bibitem[Blain et al.(2003)]{2003MNRAS.338..733B} Blain, A.~W., Barnard,  V.~E., \& Chapman, S.~C.\ 2003, MNRAS, 338, 733 
\bibitem{} Carilli, C.~L., Menten, K.~M., \& Yun, M.~S.\ 1999, \apjl, 521, L25 
\bibitem{} Carilli, C.~L., et al.\ 2002, \aj, 123, 1838 
\bibitem[Carilli \& Walter(2013)]{2013ARA&A..51..105C} Carilli, C.~L., \& Walter, F.\ 2013, \araa, 51, 105 
\bibitem[Carilli et al.(2013)]{2013ApJ...763..120C} Carilli, C.~L., Riechers, D., Walter, F., et al.\ 2013, \apj, 763, 120 
\bibitem[Carniani et al.(2013)]{2013arXiv1308.5113C} Carniani, S., Marconi, 
A., Biggs, A., et al.\ 2013, arXiv:1308.5113 
\bibitem[Condon(1992)]{1992ARA&A..30..575C} Condon, J.~J.\ 1992, \araa, 30, 575 
\bibitem[Downes \& Solomon(1998)]{1998ApJ...507..615D} Downes, D., \& Solomon, P.~M.\ 1998, \apj, 507, 615 
\bibitem[Dunne et al.(2000)]{2000MNRAS.315..115D} Dunne, L., Eales, S.,  Edmunds, M., et al.\ 2000, \mnras, 315, 115 
\bibitem[]{2008MNRAS.384..449F} Feroz F., Hobson M.~P., 2008, MNRAS, 384, 449
\bibitem[]{2009MNRAS.398.1601F} Feroz F., Hobson M.~P., Bridges M., 2009, MNRAS, 398, 1601
\bibitem[Galametz et al.(2012)]{2012MNRAS.425..763G} Galametz, M., Kennicutt, R.~C., Albrecht, M., et al.\ 2012, \mnras, 425, 763 
\bibitem[]{} Guilloteau, S., Omont, A., McMahon, R.~G., Cox, P., \& Petitjean, P.\ 1997, \aap, 328, L1 
\bibitem[]{} Hu, E., McMahon, R.G., Egami, E. 1996, \apjl, 459, L53
\bibitem{} Iono, D., et al.\ 2006, \apjl, 645, L97
\bibitem[Ivison et al.(2010)]{2010A&A...518L..35I} Ivison, R.~J., Swinbank, A.~M., Swinyard, B., et al.\ 2010, \aap, 518, L35 
\bibitem[Ivison et al.(2011)]{2011MNRAS.412.1913I} Ivison, R.~J.,  Papadopoulos, P.~P., Smail, I., et al.\ 2011, \mnras, 412, 1913 
\bibitem[Irwin et al.(1991)]{1991ASPC...21..117I} Irwin, M., McMahon, R.~G., \& Hazard, C.\ 1991, The Space Distribution of Quasars, 21, 117 
\bibitem{} Kennicutt, R. C. 1998, ARA\&A, 36, 189
\bibitem{2013MNRAS.430.2454L} Lentati, L., Carilli, C., Alexander, P., et al.\ 2013, \mnras, 430, 2454 
\bibitem{} Maiolino, R., et al.\ 2005, \aap, 440, L51 
\bibitem{} Magnelli, B., Lutz, D., Santini, P., et al.\ 2012, A\&A, 539, A155 
\bibitem[Ohta et al.(1996)]{1996Natur.382..426O} Ohta, K., Yamada, T., Nakanishi, K., et al.\ 1996, \nat, 382, 426 
\bibitem{1996Natur.382..428O} Omont, A., Petitjean, P.,  Guilloteau, S., McMahon, R.~G., Solomon, P.~M., \& P{\'e}contal, E.\ 1996, \nat, 382, 428 
\bibitem[Papadopoulos et al.(2012)]{2012ApJ...751...10P} Papadopoulos, P.~P., van der Werf, P., Xilouris, E., Isaak, K.~G., \& Gao, Y.\ 2012, \apj, 751, 10 
\bibitem[Priddey \& McMahon(2001)]{2001MNRAS.324L..17P} Priddey, R.~S., \& McMahon, R.~G.\ 2001, \mnras, 324, L17 
\bibitem{} Riechers, D. A., et al. 2006, ApJ, 650, 604 
\bibitem{} Riechers, D.~A., Walter, F., Carilli, C.~L., Bertoldi, F., \& Momjian, E.\ 2008, \apjl, 686, L9 
\bibitem[Riechers et al.(2011)]{2011ApJ...733L..11R} Riechers, D.~A., Carilli, L.~C., Walter, F., et al.\ 2011, \apjl, 733, L11 
\bibitem[Salom{\'e} et 
al.(2012)]{2012A&A...545A..57S} Salom{\'e}, P., Gu{\'e}lin, M., Downes, D., et al.\ 2012, \aap, 545, A57 
\bibitem{2004AIPC..735..395S} Skilling J., 2004, AIPC, 735, 395
\bibitem{} Solomon, P.~M., Downes, D., \& Radford, S.~J.~E.\ 1992, \apjl, 398, L29 
\bibitem{} Spergel, D. N. et al. 2007, ApJS, 170, 377 
\bibitem{} Stacey, G.~J., Geis, N., Genzel, R., Lugten, J.~B., Poglitsch, A., Sternberg, A., \& Townes, C.~H.\ 1991, ApJ, 373, 423 
\bibitem{2010ApJ...724..957S} Stacey, G.~J., Hailey-Dunsheath, S., Ferkinhoff, C., Nikola, T., Parshley, S.~C., Benford,  D.~J., Staguhn, J.~G., \& Fiolet, N.\ 2010, \apj, 724, 957 
\bibitem[Storrie-Lombardi et al.(1996)]{1996ApJ...468..121S}  Storrie-Lombardi, L.~J., McMahon, R.~G., Irwin, M.~J., \& Hazard, C.\ 1996, \apj, 468, 121 
\bibitem[Thomson et al.(2012)]{2012MNRAS.425.2203T} Thomson, A.~P., Ivison,  R.~J., Smail, I., et al.\ 2012, \mnras, 425, 2203 
\bibitem[Venemans et al.(2012)]{2012ApJ...751L..25V} Venemans, B.~P.,  McMahon, R.~G., Walter, F., et al.\ 2012, \apjl, 751, L25 
\bibitem{2010A&A...519L...1W} Wagg, J., Carilli, C.~L., Wilner, D.~J., Cox, P., De Breuck, C., Menten, K., Riechers, D.~A., \& Walter, F.\ 2010, \aap, 519, L1 
\bibitem[Wagg et al.(2012)]{2012ApJ...752L..30W} Wagg, J., Wiklind, T., Carilli, C.~L., et al.\ 2012, \apjl, 752, L30 
\bibitem{} Wang, R., Wagg, J., Carilli, C.~L., et al.\ 2008, AJ, 135, 1201 
\bibitem[Wang et al.(2013)]{2013ApJ...773...44W} Wang, R., Wagg, J., Carilli, C.~L., et al.\ 2013, \apj, 773, 44 
\bibitem{2000ApJ...528..171Y} Yun, M.~S., Carilli, C.~L., Kawabe, R., Tutui, Y., Kohno, K., \& Ohta, K.\ 2000, \apj, 528, 171 
\bibitem{yc2002} Yun, M.S., \& Carilli, C.L. 2002, ApJ, 568, 88


\end{thebibliography}
\end{document}